\documentclass[prl,twocolumn,floatfix]{revtex4}

\usepackage{graphicx}
\usepackage{textcomp}

\begin{document}

\title{Nonlinearities and Parametric Amplification in Superconducting Coplanar Waveguide Resonators}

\author{Erik A. Thol\'en}
\affiliation{Nanostructure Physics, Royal Institute of Technology, 10691 Stockholm, Sweden}

\author{Adem Erg\"ul}
\affiliation{Nanostructure Physics, Royal Institute of Technology, 10691 Stockholm, Sweden}

\author{Evelyn M. Doherty}
\affiliation{Nanostructure Physics, Royal Institute of Technology, 10691 Stockholm, Sweden}

\author{Frank M. Weber}
\affiliation{Nanostructure Physics, Royal Institute of Technology, 10691 Stockholm, Sweden}

\author{Fabien Gr\'egis}
\affiliation{Nanostructure Physics, Royal Institute of Technology, 10691 Stockholm, Sweden}

\author{David B. Haviland}
\email[]{haviland@kth.se}
\affiliation{Nanostructure Physics, Royal Institute of Technology, 10691 Stockholm, Sweden}

\date{\today}

\begin{abstract}
Experimental investigations of the nonlinear properties of superconducting niobium coplanar waveguide resonators are reported. The nonlinearity due to a current dependent kinetic inductance of the center conductor is strong enough to realize bifurcation of the nonlinear oscillator. When driven with two frequencies near the threshold for bifurcation, parametric amplification with a gain of +22.4~dB is observed.
\end{abstract}

\maketitle

The mesoscopic physics community has recently shown a strong interest in superconducting coplanar waveguide (CPW) resonators following ground-breaking experiments which demonstrate circuit quantum electrodynamics (QED) \cite{wallraff:cavity:04}. Fabricated with standard circuit lithography techniques, these one dimensional resonators are particularly appropriate for experiments on strong-coupling QED due to the high Q-factor and small transverse dimensions in comparison with the resonant electromagnetic wave length \cite{blais:cavity-QED:04}. Experiments to date have used such resonators in the linear regime, probing the entanglement of the quantum states of Josephson junction circuits with the harmonic oscillator states of the resonator \cite{wallraff:cavity:04}. However, nonlinear properties of the resonators make them also suitable for use as bifurcation amplifiers \cite{siddiqi:bifurcation:04} and parametric amplifiers \cite{yurke:joseph-par-amp:89}. Here we present experimental results where the intrinsic nonlinearity of the resonator is used to create strong parametric amplification.

Current-induced magnetic field penetration into superconducting thin films gives rise to a nonlinear inductance and resistance of the film \cite{dahm:intermodulation:96}. These nonlinearities cause mixing of signals and generation of new tones, which give an undesired intermodulation distortion in the context of microwave filter design \cite{randy:cellphones:04}. However these same nonlinearities can also be used to realize a desirable effect known as parametric amplification. When the resonator is operated as a parametric amplifier, two tones within the bandwidth of the resonator are applied to the input. The ``pump'' at frequency $f_p$ and the ``signal'' at frequency $f_s$ will be mixed by the nonlinearity, and new ``idler'' tones at \mbox{$2f_p-f_s$} and \mbox{$2f_s-f_p$} will appear \cite{monaco:nb-intermodualtion:00}. In the presence of a strong pump, the amplitude of the signal and idler tones can become larger than the applied signal amplitude, resulting in signal gain, or parametric amplification. The parametric amplifier is phase sensitive, amplifying one quadrature of the signal, while deamplifying the other quadrature. This amplification and deamplification applies to all ``signals'' in the resonator bandwidth, including noise. The deamplification of noise is called noise squeezing. Not only thermal noise, but also quantum zero-point fluctuations can be deamplified. The later can occur at low temperatures such that $hf_0 \ll k_B T$, resulting in squeezed quantum states in the resonator. Parametric amplification and deamplification of microwave signals with superconductors has previously been realized with a Josephson junction as the nonlinear oscillator \cite{yurke:joseph-par-amp:89} and squeezing of quantum noise was demonstrated in an experiment where the temperature of the Josephson junction was kept below the zero-point energy of the Josephson plasma oscillation \cite{movshovivh:zero-point-squeezing:90}. 

Yurke and Buks \cite{yurke:kerr-performance:06} have recently given a theory of parametric amplification in superconducting resonators. The current-dependent kinetic inductance was modeled as a Kerr nonlinearity and nonlinear dissipation due to a two-photon loss mechanism was included in the model. Kerr nonlinearity in superconducting resonators has been reported in the experimental literature \cite{chin:nonlinear_nb:92, oates:transmission-line-model:93, ma:power_ybco:97, boaknin:JJresonator:07}. Often however, nonlinear effects result from weak links which cause local phase slips \cite{hedges:weaklink:90, portis:power_switching:91} and heating/cooling oscillations in the driven resonator \cite{abdo:nonlinear:06}. Such anharmonic behavior can be used to accomplish parametric amplification \cite {abdo:intermod-gain-nbn:06}, however it is unclear whether this more complicated type of nonlinear oscillator is appropriate for quantum noise squeezing. The resonators reported on here are well described by the simple Kerr nonlinearity and demonstrate high parametric gain.

\begin{figure}
	\includegraphics{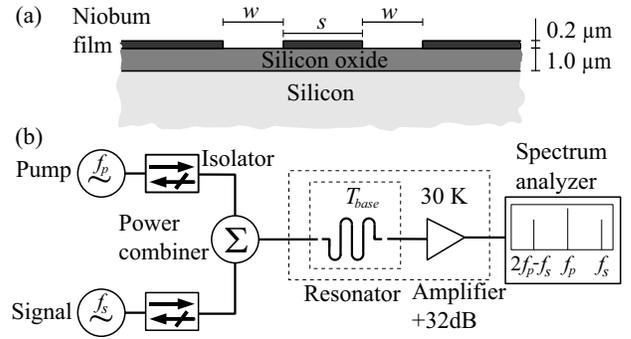}
	\caption{(a) Cross-section of the resonator. (b) Schematic of the measurement.}
	\label{fig:schematic}
\end{figure}

CPW resonators with cross-section as shown in fig. 1(a) were fabricated from 200 nm thick niobium films sputtered onto $\mathrm{SiO_2/Si}$ wafers. E-beam lithography and reactive ion etching with $\mathrm{BCl_3}$ was used to define the resonator and input/output structures. Both ends of the resonator are capacitively coupled to transmission lines making it a two-port device which is measured in transmission. Several samples with different dimensions $s$ and $w$, and different coupling capacitors have been studied. Two samples discussed here are: sample~1 with \mbox{$s=10$~\textmu m}, \mbox{$w=4$~\textmu m}, length 22.9~mm, fundamental resonance frequency \mbox{$f_0=3.1$~GHz}, \mbox{Q-factor~=~55000} and insertion loss at resonance of $S_{21}(f_0)=$ --11.4~dB; and sample~2 with \mbox{$s = 1$~\textmu m}, \mbox{$w = 1$~\textmu m}, length 13.2~mm, \mbox{$f_0=5.8$~GHz}, \mbox{$Q=36000$} and $S_{21}(f_0)=$ --11.2~dB. Both resonators are undercoupled, i.e., resistive losses in the center strip are larger than radiative losses through the coupling capacitors.

The chip containing the sample was mounted and wire-bonded to a printed circuit (PC) board which was placed in a copper box. MMCX coaxial connectors were used to couple cryogenic coaxial cables to the PC board. The sample box was cooled in a dip-stick style dilution refrigerator capable of base temperature $T_{base}=25$~mK. A schematic of the measurement setup is shown in fig 1(b). Reported power levels refer to the input or output of the sample box, where the frequency dependent cable losses and amplifier gain have been calibrated away.

\begin{figure}
	\includegraphics{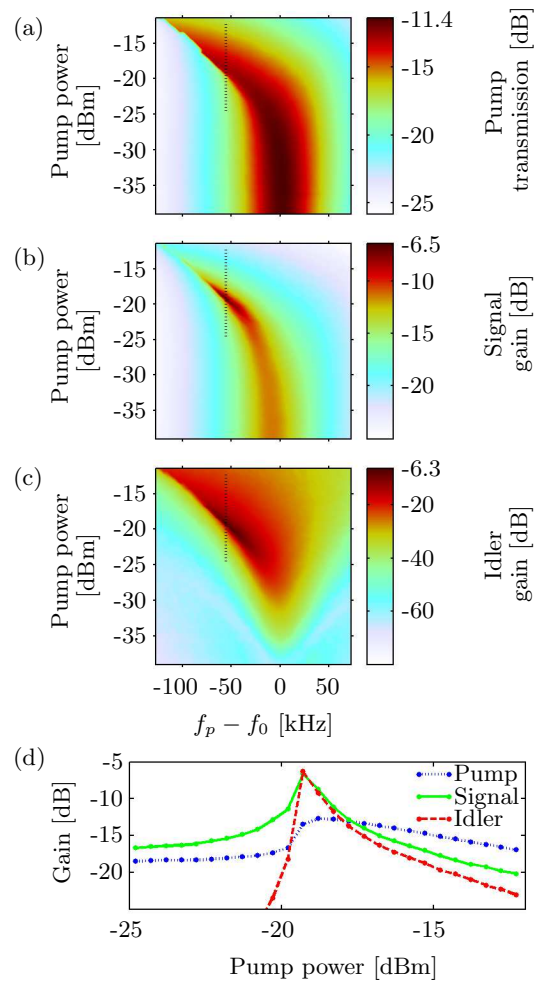}
	\caption{(Color online) Output power at (a) $f_p$, (b) $f_s$ and (c) \mbox{$2f_p-f_s$} plotted by shade versus pump power and frequency for sample 1. The signal and idler powers are maximum around the point where pump bifurcates. (d) All signals versus pump power at the frequency cut indicated by the dotted lines in (a)--(c).}
	\label{fig:evelyn_sample}
\end{figure}

In Figure~2(a)-(c) the shade (color online) shows the pump, signal and idler output power levels respectively, plotted versus the frequency and the pump power. Measurements were made sweeping $f_p$ and $f_s$ from low to high frequencies, keeping $\Delta f = f_s-f_p = 10$~kHz constant, and stepping the pump power with constant signal power --40~dBm. Fig.~2(a) shows a Lorentzian line shape at low powers, characteristic for a linear oscillator. The transmission \mbox{($S_{21}$)} on resonance is --11.4~dB. At higher power bending of the resonance peak to lower frequency is observed as expected for nonlinearity due to current dependent kinetic inductance. Figure~2(b) shows the signal gain, i.e., the output power at $f_s$ divided by input power at $f_s$. Like the pump peak at low power, the signal also shows --11.4~dB loss on resonance, and bending to lower frequency with increasing pump power. Above a critical pump power, the resonator develops a region of bi-stability, where all measured properties depend on the direction of sweep (see appendix). At the point where the pump response becomes unstable, the signal output power increases by about +5~dB. Thus parametric amplification is observed, but there is no overall gain due to the low transmission on resonance. In figure~2(c) we see the idler gain (output power at \mbox{$2f_p-f_s$} divided by input power at $f_s$) which reaches about the same level as the signal gain. Figure 2(d) shows a cut taken along the dotted lines in figs.~2(a)-(c). Here we clearly see how the the signal and idler gains are maximal at the pump power where the resonator abruptly changes from low amplitude to high amplitude oscillations.

\begin{figure}
	\includegraphics{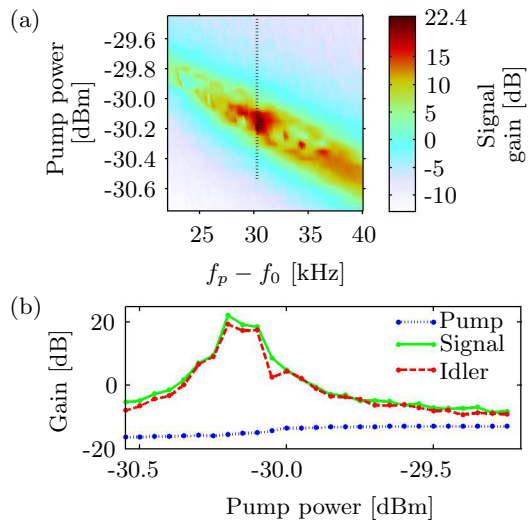}
	\caption{(Color online) (a) Signal gain for sample 2 in a narrow region around the optimal point. The fluctuations in gain are due to problems with self-heating and unstable temperature. (b) All signals versus pump power at the frequency cut indicated by the dotted lines in (a).}
	\label{fig:adem_sample}
\end{figure}

Figure~3(a) shows the signal gain for sample~2 measured with $\Delta f = 5$~kHz and signal power \mbox{--85~dBm}. The smaller cross-section of this resonator leads to enhanced nonlinear inductance, and consequently the onset of nonlinear behavior was observed at much lower power. In fig.~3 we show only a zoomed-in region around the onset of the instability, with more dense sampling in power and frequency. As predicted by theory \cite{yurke:kerr-performance:06} the gain is very sharply peaked at the critical pump power. We observe as high as +22.4~dB referred to the input of the sample box. Figure 3(b) shows a cut along the dashed line of figure 3(a). Here again the idler gain reaches the same level as the signal gain.

The nonlinear behavior can be modeled by expanding the distributed inductance and resistance of the transmission line to lowest order in the current. We have numerically solved the nonlinear transmission line equations, including the capacitive termination, using the method of harmonic balance \cite{oates:transmission-line-model:93}. Excellent fits of the model to the measured transmission versus drive power were achieved over a wide range of drive power (see appendix). The current dependent inductance and resistance per unit legnth, and the peak current in the resonator at the onset of bifurcation $I_b$ were found to be: Sample 1, $L(i) = 330 + 0.018 i^2$ nH/m, $R(i) = 86 + 11 i^2 \mathrm{~m\Omega /m}$, $I_b = 84$ mA; Sample 2, $L(i) = 337 + 0.027 i^2$ nH/m, $R(i) = 194 + 30 i^2 \mathrm{~m\Omega /m}$, $I_b = 26$ mA, where $i=I/I_b$.

We have also measured the phase-dependence of the amplification, verifying that we can indeed observe deamplification. Mixing the output of the resonator with the phase-shifted pump signal, we found a phase shift of $\pi /2$ between maximum amplification and maximum deamplification \cite{yurke:joseph-par-amp:89}. One problem that we encountered with these parametric amplifiers was heating of the sample due to the high pump power and internal dissipation in the resonator. We could verify the heating effect because at temperatures below $\approx 1$ K, all resonators exhibited a small \textit{increase} in the resonance frequency with increasing temperature. From this measured temperature dependence, we can determine the temperature of the resonator when self-heated by the continuous pump, which was about 900~mK for Sample 1 and 450~mK for sample 2. The lower temperature of operation for sample 2 is a result of lower pump power, due to a stronger nonlinear inductance. With further optimization of the sample design, we are confident that parametric amplification can be achieved in the interesting regime where $h f_0 < k_B T$.

Summarizing our main results: Current dependent kinetic inductance of a superconducting CPW resonator was used to realize strong parametric amplification. The gain was sharply peaked at the critical pump power. A maximum gain of +22.4~dB was achieved. 

This work was supported by the Swedish SSF, VR and the K. A. Wallenberg Foundation. Niobium films were produced by by A. Pavolotsky at Chalmers. We acknoledge helpful conversations with A. Wallraff, R. Schoelkopf and M. Devoret.

\newpage
\onecolumngrid
\appendix
\section{Appendix: Model description and fits to measurement data}

\begin{figure}[h]
\centering
\includegraphics[width=0.65\textwidth]{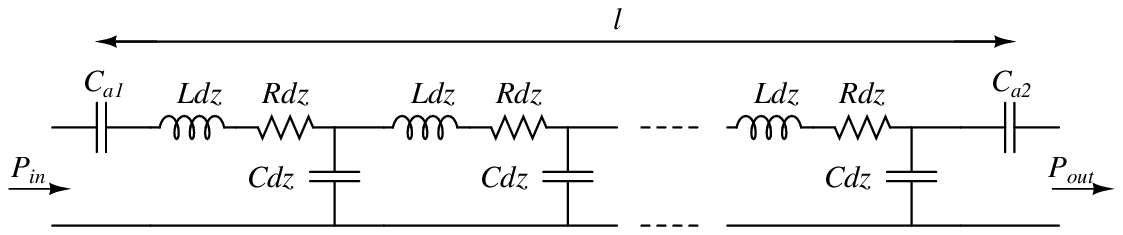}
\end{figure}

FIG. 4: The resonator is modelled as a continous transmission line with length $l$ between two discrete capacitors $C_{a1}$ and $C_{a2}$. Outside the capacitors are $50~\mathrm{\Omega}$ transmission lines.

\vspace*{0.6cm}
\begin{minipage}{0.45\linewidth}
	\includegraphics{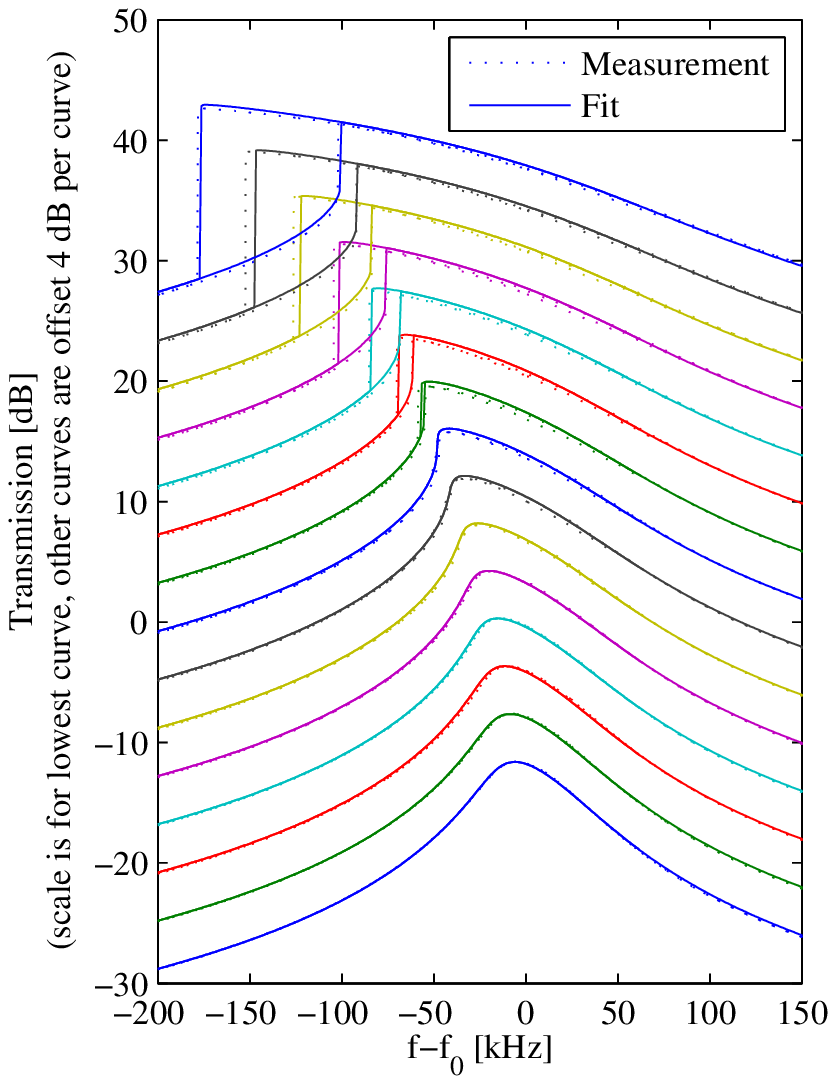}
\end{minipage}
\hfill
\begin{minipage}{0.49\textwidth}
FIG. 5: The transmission is calculated by integrating the telegrapher equations
\begin{eqnarray*}
	\frac{\partial V}{\partial z} & = & -\frac{\partial}{\partial t}(LI) - IR \\
	\frac{\partial I}{\partial z} & = & -\frac{\partial}{\partial t}(CV)
\end{eqnarray*}
over the length of the resonator while satisfying the boundary conditions given by the capacitors and transmission lines on both sides. The model includes current dependent (i.e. nonlinear) inductance and resistance (see table 1). For details about the model see reference [11]. 



The figure shows our results for sample 1 but the same procedure was also performed for samle 2. The lowest curve shows the transmission spectrum ($P_{out}/P_{in}$) of the resonator at a relatively low power \mbox{(-27 dBm)} where there are no nonlinear effects. For each of the curves above it the input power has been increased by 1 dBm. For clarity each curve has been offset by 4 dB. Above a critical power there is a region of bistability where the oscillation can have two different amplitudes depending on the direction of the frequency sweep. The intermodulation products are also affected by this hysteresis. A peak of the gain is observed at the unstable points for both upward and downward sweeps of frequency.
\end{minipage}

\vspace*{0.6cm}
\begin{minipage}{0.55\linewidth}
	\begin{tabular}{|l|l|l|r|r|r|}
	\hline
	Parameter               & Symb.   	& Note       & Sample 1          & Sample 2 	  & Unit \\ \hline \hline
	Port capacitance        & $C_a$  	& Fitted     & $2.08$            & $1.66$            & fF \\ \hline
	Inductance              & $L(i)$ 	& Fitted     & $330 + 0.018 i^2$ & $337 + 0.027 i^2$ & nH/m \\ \hline
	Resistance              & $R(i)$ 	& Fitted     & $86  + 11 i^2$    & $194  + 30 i^2$   & $\mathrm{m\Omega /m}$ \\ \hline
	Bifurcation power       & $P_b$  	& Measured   & $-20.0$           & $-30.2$              & dBm \\ \hline
	Bifurcation current     & $I_b$  	& Calculated & $84$              & $26$              & mA \\ \hline
	Capacitance             & $C$    	& Calculated & $151$             & $125$	          & pF/m \\ \hline
	Resonator length        & $l$    	& Designed   & $22.9$		 & $13.2$	          & mm \\ \hline
	Film thickness	        & $t$    	& Designed   & $0.2$	         & $0.2$	          & \textmu m \\ \hline
	Strip width	        & $s$    	& Designed   & $10$	      	 & $1$		  & \textmu m \\ \hline
	Gap width	        & $w$		& Designed   & $4$		 & $1$	          & \textmu m \\ \hline
	\end{tabular}
\end{minipage}
\hfill
\begin{minipage}{0.33\linewidth}
TABLE 1: The parameters used in the model. The current $i=I/I_b$ is normalized to $I_b$ which is the maximal current in the resonator at the bifurcation point.

\end{minipage}









\begin{thebibliography}{17}
\expandafter\ifx\csname natexlab\endcsname\relax\def\natexlab#1{#1}\fi
\expandafter\ifx\csname bibnamefont\endcsname\relax
  \def\bibnamefont#1{#1}\fi
\expandafter\ifx\csname bibfnamefont\endcsname\relax
  \def\bibfnamefont#1{#1}\fi
\expandafter\ifx\csname citenamefont\endcsname\relax
  \def\citenamefont#1{#1}\fi
\expandafter\ifx\csname url\endcsname\relax
  \def\url#1{\texttt{#1}}\fi
\expandafter\ifx\csname urlprefix\endcsname\relax\def\urlprefix{URL }\fi
\providecommand{\bibinfo}[2]{#2}
\providecommand{\eprint}[2][]{\url{#2}}

\bibitem[{\citenamefont{Wallraff et~al.}(2004)\citenamefont{Wallraff, Schuster,
  Blais, Frunzio, Huang, Majer, Kumar, Girvin, and
  Schoelkopf}}]{wallraff:cavity:04}
\bibinfo{author}{\bibfnamefont{A.}~\bibnamefont{Wallraff}},
  \bibinfo{author}{\bibfnamefont{D.~I.} \bibnamefont{Schuster}},
  \bibinfo{author}{\bibfnamefont{A.}~\bibnamefont{Blais}},
  \bibinfo{author}{\bibfnamefont{L.}~\bibnamefont{Frunzio}},
  \bibinfo{author}{\bibfnamefont{R.-S.} \bibnamefont{Huang}},
  \bibinfo{author}{\bibfnamefont{J.}~\bibnamefont{Majer}},
  \bibinfo{author}{\bibfnamefont{S.}~\bibnamefont{Kumar}},
  \bibinfo{author}{\bibfnamefont{S.~M.} \bibnamefont{Girvin}},
  \bibnamefont{and} \bibinfo{author}{\bibfnamefont{R.~J.}
  \bibnamefont{Schoelkopf}}, \bibinfo{journal}{Nature}
  \textbf{\bibinfo{volume}{431}}, \bibinfo{pages}{162} (\bibinfo{year}{2004}).

\bibitem[{\citenamefont{Blais et~al.}(2004)\citenamefont{Blais, Huang,
  Wallraff, Girvin, and Schoelkopf}}]{blais:cavity-QED:04}
\bibinfo{author}{\bibfnamefont{A.}~\bibnamefont{Blais}},
  \bibinfo{author}{\bibfnamefont{R.~S.} \bibnamefont{Huang}},
  \bibinfo{author}{\bibfnamefont{A.}~\bibnamefont{Wallraff}},
  \bibinfo{author}{\bibfnamefont{S.~M.} \bibnamefont{Girvin}},
  \bibnamefont{and} \bibinfo{author}{\bibfnamefont{R.~J.}
  \bibnamefont{Schoelkopf}}, \bibinfo{journal}{Phys. Rev. A}
  \textbf{\bibinfo{volume}{69}}, \bibinfo{pages}{062320}
  (\bibinfo{year}{2004}).

\bibitem[{\citenamefont{Siddiqi et~al.}(2004)\citenamefont{Siddiqi, Vijay,
  Pierre, Wilson, Metcalfe, Rigetti, Frunzio, and
  Devoret}}]{siddiqi:bifurcation:04}
\bibinfo{author}{\bibfnamefont{I.}~\bibnamefont{Siddiqi}},
  \bibinfo{author}{\bibfnamefont{R.}~\bibnamefont{Vijay}},
  \bibinfo{author}{\bibfnamefont{F.}~\bibnamefont{Pierre}},
  \bibinfo{author}{\bibfnamefont{C.~M.} \bibnamefont{Wilson}},
  \bibinfo{author}{\bibfnamefont{M.}~\bibnamefont{Metcalfe}},
  \bibinfo{author}{\bibfnamefont{C.}~\bibnamefont{Rigetti}},
  \bibinfo{author}{\bibfnamefont{L.}~\bibnamefont{Frunzio}}, \bibnamefont{and}
  \bibinfo{author}{\bibfnamefont{M.~H.} \bibnamefont{Devoret}},
  \bibinfo{journal}{Phys. Rev. Lett.} \textbf{\bibinfo{volume}{93}},
  \bibinfo{pages}{207002} (\bibinfo{year}{2004}).

\bibitem[{\citenamefont{Yurke et~al.}(1989)\citenamefont{Yurke, Corruccini,
  Kaminsky, Rupp, Smith, Silver, Simon, and
  Whittaker}}]{yurke:joseph-par-amp:89}
\bibinfo{author}{\bibfnamefont{B.}~\bibnamefont{Yurke}},
  \bibinfo{author}{\bibfnamefont{L.~R.} \bibnamefont{Corruccini}},
  \bibinfo{author}{\bibfnamefont{P.~G.} \bibnamefont{Kaminsky}},
  \bibinfo{author}{\bibfnamefont{L.~W.} \bibnamefont{Rupp}},
  \bibinfo{author}{\bibfnamefont{A.~D.} \bibnamefont{Smith}},
  \bibinfo{author}{\bibfnamefont{A.~H.} \bibnamefont{Silver}},
  \bibinfo{author}{\bibfnamefont{R.~W.} \bibnamefont{Simon}}, \bibnamefont{and}
  \bibinfo{author}{\bibfnamefont{E.~A.} \bibnamefont{Whittaker}},
  \bibinfo{journal}{Phys. Rev. A} \textbf{\bibinfo{volume}{39}},
  \bibinfo{pages}{2519} (\bibinfo{year}{1989}).

\bibitem[{\citenamefont{Dahm and Scalapino}(1996)}]{dahm:intermodulation:96}
\bibinfo{author}{\bibfnamefont{T.}~\bibnamefont{Dahm}} \bibnamefont{and}
  \bibinfo{author}{\bibfnamefont{D.~J.} \bibnamefont{Scalapino}},
  \bibinfo{journal}{J. Appl. Phys} \textbf{\bibinfo{volume}{81}},
  \bibinfo{pages}{2002} (\bibinfo{year}{1996}).

\bibitem[{\citenamefont{Simon et~al.}(2004)\citenamefont{Simon, Hammond,
  Berkowitz, and Willemsen}}]{randy:cellphones:04}
\bibinfo{author}{\bibfnamefont{R.~W.} \bibnamefont{Simon}},
  \bibinfo{author}{\bibfnamefont{R.~B.} \bibnamefont{Hammond}},
  \bibinfo{author}{\bibfnamefont{S.~J.} \bibnamefont{Berkowitz}},
  \bibnamefont{and} \bibinfo{author}{\bibfnamefont{B.~A.}
  \bibnamefont{Willemsen}}, in \emph{\bibinfo{booktitle}{Proceedings of the
  IEEE}} (\bibinfo{year}{2004}), vol.~\bibinfo{volume}{92}, p.
  \bibinfo{pages}{1585}.

\bibitem[{\citenamefont{Monaco et~al.}(2000)\citenamefont{Monaco, Andreone, and
  Palomba}}]{monaco:nb-intermodualtion:00}
\bibinfo{author}{\bibfnamefont{R.}~\bibnamefont{Monaco}},
  \bibinfo{author}{\bibfnamefont{A.}~\bibnamefont{Andreone}}, \bibnamefont{and}
  \bibinfo{author}{\bibfnamefont{F.}~\bibnamefont{Palomba}},
  \bibinfo{journal}{J. Appl. Phys.} \textbf{\bibinfo{volume}{88}},
  \bibinfo{pages}{2898} (\bibinfo{year}{2000}).

\bibitem[{\citenamefont{Movshovich et~al.}(1990)\citenamefont{Movshovich,
  Yurke, Kaminsky, Smith, Silver, Simon, and
  Schneider}}]{movshovivh:zero-point-squeezing:90}
\bibinfo{author}{\bibfnamefont{R.}~\bibnamefont{Movshovich}},
  \bibinfo{author}{\bibfnamefont{B.}~\bibnamefont{Yurke}},
  \bibinfo{author}{\bibfnamefont{P.~G.} \bibnamefont{Kaminsky}},
  \bibinfo{author}{\bibfnamefont{A.~D.} \bibnamefont{Smith}},
  \bibinfo{author}{\bibfnamefont{A.~H.} \bibnamefont{Silver}},
  \bibinfo{author}{\bibfnamefont{R.~W.} \bibnamefont{Simon}}, \bibnamefont{and}
  \bibinfo{author}{\bibfnamefont{M.~V.} \bibnamefont{Schneider}},
  \bibinfo{journal}{Phys. Rev. Lett.} \textbf{\bibinfo{volume}{65}},
  \bibinfo{pages}{1419} (\bibinfo{year}{1990}).

\bibitem[{\citenamefont{Yurke and Buks}(2006)}]{yurke:kerr-performance:06}
\bibinfo{author}{\bibfnamefont{B.}~\bibnamefont{Yurke}} \bibnamefont{and}
  \bibinfo{author}{\bibfnamefont{E.}~\bibnamefont{Buks}}, \bibinfo{journal}{J.
  Lightw. Technol.} \textbf{\bibinfo{volume}{24}}, \bibinfo{pages}{5054}
  (\bibinfo{year}{2006}).

\bibitem[{\citenamefont{Chin et~al.}(1992)\citenamefont{Chin, Oates,
  Dresselhaus, and Dresselhaus}}]{chin:nonlinear_nb:92}
\bibinfo{author}{\bibfnamefont{C.~C.} \bibnamefont{Chin}},
  \bibinfo{author}{\bibfnamefont{D.~E.} \bibnamefont{Oates}},
  \bibinfo{author}{\bibfnamefont{G.}~\bibnamefont{Dresselhaus}},
  \bibnamefont{and} \bibinfo{author}{\bibfnamefont{M.~S.}
  \bibnamefont{Dresselhaus}}, \bibinfo{journal}{Phys. Rev. B}
  \textbf{\bibinfo{volume}{45}}, \bibinfo{pages}{4788} (\bibinfo{year}{1992}).

\bibitem[{\citenamefont{Oates et~al.}(1993)\citenamefont{Oates, Shin, Oates,
  Tsuk, and Nguyen}}]{oates:transmission-line-model:93}
\bibinfo{author}{\bibfnamefont{J.~H.} \bibnamefont{Oates}},
  \bibinfo{author}{\bibfnamefont{R.~T.} \bibnamefont{Shin}},
  \bibinfo{author}{\bibfnamefont{D.~E.} \bibnamefont{Oates}},
  \bibinfo{author}{\bibfnamefont{M.~J.} \bibnamefont{Tsuk}}, \bibnamefont{and}
  \bibinfo{author}{\bibfnamefont{P.~P.} \bibnamefont{Nguyen}},
  \bibinfo{journal}{IEEE Trans. Appl. Supercond.} \textbf{\bibinfo{volume}{3}},
  \bibinfo{pages}{17} (\bibinfo{year}{1993}), \bibinfo{note}{part 4}.

\bibitem[{\citenamefont{Ma et~al.}(1997)\citenamefont{Ma, de~Obaldia, Hampel,
  Polakos, Mankiewich, Batlogg, Prusseit, Kinder, Anderson, Oates
  et~al.}}]{ma:power_ybco:97}
\bibinfo{author}{\bibfnamefont{Z.}~\bibnamefont{Ma}},
  \bibinfo{author}{\bibfnamefont{E.}~\bibnamefont{de~Obaldia}},
  \bibinfo{author}{\bibfnamefont{G.}~\bibnamefont{Hampel}},
  \bibinfo{author}{\bibfnamefont{P.}~\bibnamefont{Polakos}},
  \bibinfo{author}{\bibfnamefont{P.}~\bibnamefont{Mankiewich}},
  \bibinfo{author}{\bibfnamefont{B.}~\bibnamefont{Batlogg}},
  \bibinfo{author}{\bibfnamefont{W.}~\bibnamefont{Prusseit}},
  \bibinfo{author}{\bibfnamefont{H.}~\bibnamefont{Kinder}},
  \bibinfo{author}{\bibfnamefont{A.}~\bibnamefont{Anderson}},
  \bibinfo{author}{\bibfnamefont{D.~E.} \bibnamefont{Oates}},
  \bibinfo{author}{\bibfnamefont{R.} \bibnamefont{Ono}},
  \bibinfo{author}{\bibfnamefont{J.} \bibnamefont{Beall}},
  \bibinfo{journal}{IEEE Trans. Appl. Supercond.}
  \textbf{\bibinfo{volume}{7}}, \bibinfo{pages}{1911} (\bibinfo{year}{1997}),
  \bibinfo{note}{part 2}.

\bibitem[{\citenamefont{Boaknin et~al.}(2007)\citenamefont{Boaknin,
  Manucharyan, Fissette, Metcalfe, Frunzio, Vijay, Siddiqi, Wallraff,
  Schoelkopf, and Devoret}}]{boaknin:JJresonator:07}
\bibinfo{author}{\bibfnamefont{E.}~\bibnamefont{Boaknin}},
  \bibinfo{author}{\bibfnamefont{V.~E.} \bibnamefont{Manucharyan}},
  \bibinfo{author}{\bibfnamefont{S.}~\bibnamefont{Fissette}},
  \bibinfo{author}{\bibfnamefont{M.}~\bibnamefont{Metcalfe}},
  \bibinfo{author}{\bibfnamefont{L.}~\bibnamefont{Frunzio}},
  \bibinfo{author}{\bibfnamefont{R.}~\bibnamefont{Vijay}},
  \bibinfo{author}{\bibfnamefont{I.}~\bibnamefont{Siddiqi}},
  \bibinfo{author}{\bibfnamefont{A.}~\bibnamefont{Wallraff}},
  \bibinfo{author}{\bibfnamefont{R.~J.} \bibnamefont{Schoelkopf}},
  \bibnamefont{and} \bibinfo{author}{\bibfnamefont{M.}~\bibnamefont{Devoret}},
  \bibinfo{journal}{arXiv:cond-mat/0702445v1}  (\bibinfo{year}{2007}).

\bibitem[{\citenamefont{Hedges et~al.}(1990)\citenamefont{Hedges, Adams,
  Nicholson, and Chew}}]{hedges:weaklink:90}
\bibinfo{author}{\bibfnamefont{S.~J.} \bibnamefont{Hedges}},
  \bibinfo{author}{\bibfnamefont{M.~J.} \bibnamefont{Adams}},
  \bibinfo{author}{\bibfnamefont{B.~F.} \bibnamefont{Nicholson}},
  \bibnamefont{and} \bibinfo{author}{\bibfnamefont{N.~G.} \bibnamefont{Chew}},
  \bibinfo{journal}{Electron. Lett.} \textbf{\bibinfo{volume}{26}},
  \bibinfo{pages}{977} (\bibinfo{year}{1990}).

\bibitem[{\citenamefont{Portis et~al.}(1991)\citenamefont{Portis, Chaloupka,
  Jeck, Piel, and Pischke}}]{portis:power_switching:91}
\bibinfo{author}{\bibfnamefont{A.~M.} \bibnamefont{Portis}},
  \bibinfo{author}{\bibfnamefont{H.}~\bibnamefont{Chaloupka}},
  \bibinfo{author}{\bibfnamefont{M.}~\bibnamefont{Jeck}},
  \bibinfo{author}{\bibfnamefont{H.}~\bibnamefont{Piel}}, \bibnamefont{and}
  \bibinfo{author}{\bibfnamefont{A.}~\bibnamefont{Pischke}},
  \bibinfo{journal}{Supercond. Sci. Technol.} \textbf{\bibinfo{volume}{4}},
  \bibinfo{pages}{436} (\bibinfo{year}{1991}).

\bibitem[{\citenamefont{Abdo et~al.}(2006{\natexlab{a}})\citenamefont{Abdo,
  Segev, Shtempluck, and Buks}}]{abdo:nonlinear:06}
\bibinfo{author}{\bibfnamefont{B.}~\bibnamefont{Abdo}},
  \bibinfo{author}{\bibfnamefont{E.}~\bibnamefont{Segev}},
  \bibinfo{author}{\bibfnamefont{O.}~\bibnamefont{Shtempluck}},
  \bibnamefont{and} \bibinfo{author}{\bibfnamefont{E.}~\bibnamefont{Buks}},
  \bibinfo{journal}{Phys. Rev. B} \textbf{\bibinfo{volume}{73}},
  \bibinfo{pages}{134513} (\bibinfo{year}{2006}{\natexlab{a}}).

\bibitem[{\citenamefont{Abdo et~al.}(2006{\natexlab{b}})\citenamefont{Abdo,
  Segev, Shtempluck, and Buks}}]{abdo:intermod-gain-nbn:06}
\bibinfo{author}{\bibfnamefont{B.}~\bibnamefont{Abdo}},
  \bibinfo{author}{\bibfnamefont{E.}~\bibnamefont{Segev}},
  \bibinfo{author}{\bibfnamefont{O.}~\bibnamefont{Shtempluck}},
  \bibnamefont{and} \bibinfo{author}{\bibfnamefont{E.}~\bibnamefont{Buks}},
  \bibinfo{journal}{Appl. Phys. Lett.} \textbf{\bibinfo{volume}{88}},
  \bibinfo{pages}{022508} (\bibinfo{year}{2006}{\natexlab{b}}).

\end{thebibliography}
\end{document}